\documentclass[12pt]{article}

\usepackage{amsmath}
\usepackage{amssymb}

\usepackage{graphicx}

\usepackage{color} 
\usepackage{ulem}
\usepackage{setspace,caption} 
\AtBeginCaption{\doublespacing}

\date{}

\def \eps {\epsilon}

\def \R {\mathbb{R}}

\def \v {\mathbf{v}}

\def \u {\mathbf{u}}
\def \x {\mathbf{x}}
\def \y {\mathbf{y}}

\def \W {\mathbf{W}}

\begin{document}
\sloppy
\begin{flushleft}
{\Large
\textbf{A biological gradient descent for prediction through a combination of STDP and homeostatic plasticity.}
}
\\
Mathieu N. Galtier$^{1}$, 
Gilles Wainrib$^{2}$
\\
\bf{1} School of Engineering and Science, Jacobs University Bremen gGmbH, College Ring 1, 28759 Bremen, Germany
\\
\bf{2} Laboratoire Analyse G\'eom\'etrie et Applications, Universit\'e Paris 13, 99 avenue Jean-Baptiste Cl\'ement, Villetaneuse, France
\end{flushleft}

\section*{Abstract}

Identifying, formalizing and combining biological mechanisms which implement known brain functions, such as prediction, is a main aspect of current research in theoretical neuroscience. In this letter, 
the mechanisms of Spike Timing Dependent Plasticity (STDP) and homeostatic plasticity, combined in an original mathematical formalism, are shown to shape recurrent neural networks into predictors. Following a rigorous mathematical treatment, we prove that they implement the online gradient descent of a distance between the network activity and its stimuli. The convergence to an equilibrium, where the network can spontaneously reproduce or predict its stimuli, does not suffer from bifurcation issues usually encountered in learning in recurrent neural networks.

\section{Introduction}
One of the main functions of the brain is prediction \cite{bar2009predictions}. This function is generally thought to rely on the idea that cortical regions learn a model of the world and simulate it to generate predictions of future events \cite{gilbert2007prospection, schacter2008episodic}. Several recent experimental findings support this view, showing in particular that the spontaneous neuronal activity after presentation of a stimulus is correlated with the evoked activity \cite{kenet2003spontaneously}, and that this similarity increases along development and learning \cite{berkes2011spontaneous}. Moreover, at the scale of neuronal networks, prediction can also be seen as a general organization principle : it has been argued \cite{rao1999predictive, clark2012whatever} that the brain would contain a hierarchy of predictive units which are able to predict their direct stimuli or inputs, through the modification of the synaptic connections.

Understanding the mechanisms and principles underlying this prediction function is a key challenge, not only from a neuroscience perspective but also for machine learning where the number of applications requiring prediction is significant.

In the field of machine learning, recurrent neural networks have been successfully proposed as candidates for these predictive units \cite{williams1989learning, williams1995gradient, pearlmutter1995gradient, jaeger2004harnessing, sussillo2009generating}. In most cases, these algorithms aim at creating a neural network that autonomously and spontaneously reproduces a given time series. The Bayesian approach is also useful in designing predictors \cite{dayan1995helmholtz, george2009towards} and has also been mapped to neural networks \cite{deneve2008bayesian,friston2010free,bitzer2012recognizing}. However, apart from a rough conceptual equivalence, this paper is devoid of Bayesian terminology and directly focuses on neural networks. In this framework, prediction is often achieved by minimizing a distance between the activity of the neural network and the target time series. Although neural networks were originally studied in a feedforward framework \cite{rosenblatt1958perceptron}, the most efficient networks 
for prediction shall involve recurrent connections giving the network some memory properties. So called gradient descent algorithms in recurrent neural network \cite{mandic2001recurrent} involve the learning of the entire connectivity matrix. They minimize the distance between a target trajectory and the trajectory of the network. On the other hand, researchers in the field of reservoir computing \cite{lukovsevivcius2009reservoir} only optimize some connections in the network whereas the others are randomly drawn and fixed. To do prediction, they minimize the ``one-step ahead'' error corresponding to the distance between the network predictions and the next time step of the target time series. Thus, these algorithms are derived to optimize an accuracy criterion, with learning rules generally favoring prediction efficiency over biological meaning.

In the field of neuroscience, these last years have seen many discoveries in the study of synaptic plasticity, in particular providing experimental evidences and possible mechanisms for two major concepts in the current biology of learning.

\noindent The first is the discovery of spike-timing dependent plasticity (STDP) ~\cite{markram1997regulation, bi1998synaptic, caporale2008spike, sjostrom:2010}. It is a temporally asymmetric form of Hebbian learning induced by temporal correlations between the spikes of pre- and post-synaptic neurons. The general principle is that if a neuron fires before (resp. after) another then the strength of the connection from the former to the latter will be increased (resp. decreased). The summation of all these modification leads to the strengthening of causality links between neurons. Although STDP is originally based on spiking network, it has several extensions or analogs for rate-based networks (those used in machine learning) \cite{kempter1999hebbian, izhikevich2003relating, 
pfister2006triplets}. The functional role of STDP is still discussed, for instance: reducing latency \cite{song2000competitive}, optimizing information transfer \cite{hennequin2010stdp}, invariant recognition \cite{sprekeler2007slowness} and even learning temporal patterns \cite{gerstner1993spikes, rao2001spike, yoshioka2007spatiotemporal} (non exhaustive list).

\noindent Second, the notion of homeostatic plasticity \cite{miller1996synaptic, abbott2000synaptic, turrigiano2004homeostatic}, including mechanisms such as synaptic scaling, has proved to be important to moderate the growth of connection strength. In contrast to previously theory-motivated normalization of the connectivity \cite{miller1994role, oja1982simplified}, there is a need of a biologically plausible means to prevent the connectivity from exploding under the influence of shaping mechanisms like Hebbian learning or STDP.

\noindent From a theoretical viewpoint, STDP and homeostatic plasticity are almost always studied independently. An extensive bottom-up numerical analysis of the combination of such learning mechanisms, done by Triesch and colleagues, has already lead to biologically relevant behaviors \cite{lazar2007, lazar2009sorn, zheng2013network}. However, the mathematical understanding of their combination in terms of functionality still stands as an undocumented challenge to researchers.

This letter aims at bridging the gap between biological mechanisms and machine learning regarding the issue of predictive neural networks. We rigorously show how a biologically-inspired learning rule, made of an original combination of STDP mechanism and homeostatic plasticity, mimics the gradient descent of a distance between the activity of the neural network and its direct stimuli. This results in capturing the underlying dynamical behavior of the stimuli into a recurrent neural network and therefore in designing a biologically plausible predictive network.

The letter is organized as follows. In section \ref{sec: gradient descent}, we construct a theoretical learning rule designed for prediction, based on an appropriate gradient descent method. Then, in section \ref{sec: biological}, we introduce a biologically inspired learning rule, combining the concepts of STDP and homeostatic plasticity, whose purpose is to mimic the theoretical learning rule. We discuss the various biological mechanisms which may be involved in this new learning rule. Finally in section \ref{sec: link}, we provide a mathematical justification of the link between the theoretical and the biologically inspired learning rule, based on the key idea that STDP can be seen as a differential operator.

\section{Theoretical learning rule for prediction}\label{sec: gradient descent}
In a machine learning approach, we introduce here a procedure to design a neural network which autonomously replays a target time series.

\subsection{Set-up}
We consider a recurrent neural network made of $n$ neurons which is exposed to a time dependent input $\u(t) \in \R^n$ of the same dimension. Our aim is to construct a learning rule which will enable the network to reproduce the input's behavior. 

{Our approach is focused on learning the underlying dynamics of the input. Therefore, we assume that $\u$ is generated by an arbitrary dynamical system:}
\begin{equation}
\dot\u = \xi(\u)
\label{eq: inputs dynamics}
\end{equation}
with $\xi$ a smooth vector field from $\R^n$ to $\R^n$. We also assume that the trajectory of the inputs or stimuli is $\tau$-periodic. The key mathematical assumptions on the input $\u$ is in fact ergodicity, but we restrict our study to periodic inputs for simplicity. In particular, periodic inputs can be constructed by the repetition of a given finite-time sample.


Although the following method virtually works with any network equations, we focus on a neural network composed of $n$ neurons and governed by
\begin{equation}
 \dot{\v} = -l \v + \W. S(\v)
\label{eq: voltage based}
\end{equation}
where $\v \in \R^{n}$ is a vector representing neuronal activity, $\W \in \R^{n \times n}$ is the connectivity matrix, $l\in \R_+$ is a decay constant and $S$ is an entry-wise sigmoid function.

\subsection{Gradient descent learning rule}
The idea behind our learning rule is to find the best connectivity matrix $\W$ which will minimize a distance between the two functions $\xi(\u)$ and $-l\u + \W.S(\u)$. In this perspective, we define the following quantity:
\begin{equation}
 H_\u := \frac{1}{2} \int_0^{\tau}\Big\|-l\u(t) + \W. S\big(\u(t)\big) - \xi\big(\u(t)\big)\Big\|^2 dt
 \label{eq: analogy relative entropy}
\end{equation}
When $H_\u = 0$, the vector fields of systems \eqref{eq: inputs dynamics} and \eqref{eq: voltage based} are equal on the trajectories of the inputs. This quantity may be viewed as a distance between the two vector fields defining the dynamics of the inputs and of the neuronal network along the trajectories of the inputs. One shall notice that it is similar to classical gradient methods~\cite{pearlmutter1995gradient, williams1995gradient, mandic2001recurrent}, except that the norms are applied to the flows of inputs and neural network instead of their activity. Thus, it focuses more specifically on the dynamical structure of the inputs. Moreover, it is possible to show, using Girsanov's Theorem, that this definition coincides with the concept of \textit{relative entropy} between two diffusion processes, namely the ones obtained by adding a standard Gaussian perturbation to both equations. Therefore, we will call $H_\u$ the relative entropy.

In order to capture the dynamics of the inputs into the network, it is natural to look for a learning rule minimizing this quantity. {To this end,} we consider the gradient of this measure with respect to the connectivity {matrix}:
\begin{equation}
   \nabla_\W H_\u = - \Big[\xi(\u)\cdot S(\u)' + l \u\cdot S(\u)' - \W. S(\u)\cdot S(\u)'\Big]
\label{eq: entropy gradient}
 \end{equation}
{where the component $i,j$ of $\mathbf x.\mathbf y' \in \R^{n \times n}$ is $\{\mathbf x\cdot \mathbf y'\}_{ij} = \int_0^{\tau} \mathbf x_i(s) \mathbf y_j(s) ds$ for $\mathbf x, \mathbf y$ functions from $[0,\tau[$ to $\R^n$}. Equivalently, these functions can be seen as semi continuous matrices in $\R^{n \times [0,\tau]}$ and $\mathbf x'$ is the transpose of $\mathbf x$.\\
Thus, an algorithm implementing the gradient descent $\dot \W = - \nabla_\W H_\u$ is a good candidate to minimize the relative entropy between inputs and spontaneous activity. Since $H_u$ is quadratic in $\W$, it follows that $\W \to H_{\u}$ is a convex function, thus excluding situations with multiple local minima. Moreover, if $S(\u)\cdot S(\u)'$ is invertible, one can compute directly the minimizing connectivity as 
\begin{equation}
\W^* = \big[\xi(\u)\cdot S(\u)' + l \u\cdot S(\u)'\big].\big[S(\u)\cdot S(\u)'\big]^{-1}
\label{eq: equilbrium connectivity}
\end{equation}

Implementing a gradient descent based on equation \eqref{eq: entropy gradient} does not immediately lead to a biologically relevant mechanism. First, it requires a direct access to the inputs $\u$, whereas synaptic plasticity mechanisms shall only rely on the network activity $\v$. Second, it is a batch learning algorithm which requires an access to the entire history of the inputs. Third, it requires the ability to compute $\xi(\u)$. Therefore, we will see in section \ref{sec: biological} how to overcome these issues combining biologically inspired synaptic plasticity mechanisms.

\subsection{Example}
\begin{figure}[ht!]
\begin{center}
 \includegraphics[width=1\textwidth]{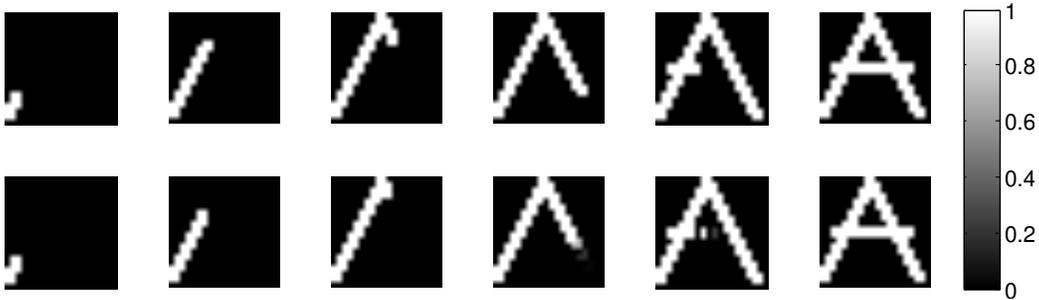}
 \end{center}
 \caption{Learning how to write the letter A. Time evolution of the input movie (top row) and of the network activity after learning (bottom row). Each pixel corresponds to a neuron.
 Parameters: Number of neurons $n=400$; $l=1$; $S(x)=\tanh(x)$.}
 \label{fig: ecrire}
\end{figure}

In order to illustrate the idea that learning rule \eqref{eq: entropy gradient} enables the network to learn dynamical features of the input, we have constructed the following experiment. We present to the network an input movie displaying sequentially the writing of the letter A (Figure \ref{fig: ecrire} - top row). To each pixel we assign one neuron, so that the input and the network share the same dimension. This input movie is repeated periodically until the connectivity matrix of the network, evolving under rule \eqref{eq: entropy gradient}, stabilizes. Then the input is turned off and we set the initial state of the network to a priming image showing the bottom left part of letter A. The network evolving without input strikingly reproduces the dynamical writing of letter A as displayed in Figure \ref{fig: ecrire}. Thus, with this example we have illustrated the ability of the learning rule we have derived from a theoretical principle to capture a dynamic input into a 
connectivity matrix.

\section{A biological learning rule}\label{sec: biological}
We now introduce a biological learning rule made of the combination of STDP and homeostatic plasticity. Later in section \ref{sec: link}, we show that this learning rule minimizes $H_\u$. Here, we first give a mathematical description of this learning rule and, second, relate the different terms to biological mechanisms.
\subsection{Mathematical description}
\begin{figure}[ht]
 \centering
 \includegraphics[width=1.\textwidth]{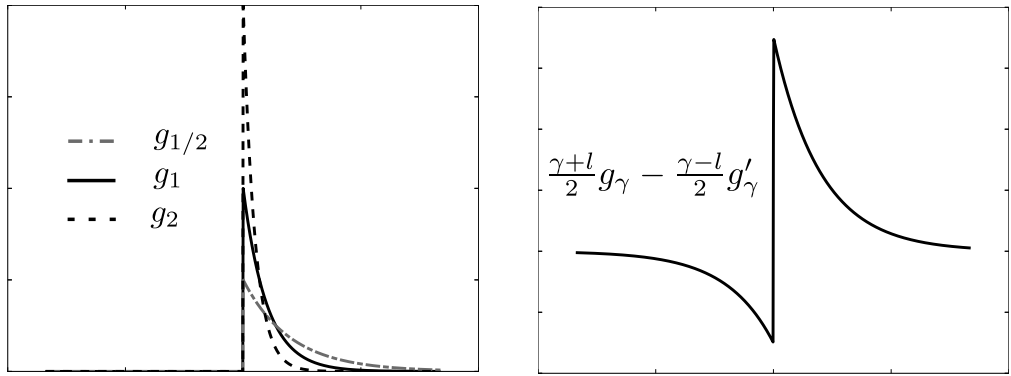}
 \caption{(left) Plots of the function $g_c$ for $c \in \{1/2, 1, 2\}$. (right) Plot of the function $\Delta_\gamma(t) + l \Sigma_\gamma(t) =  \frac{\gamma + l}{2}g_\gamma(t) - \frac{\gamma - l}{2}g_\gamma(-t)$ for $\gamma = 0.5$ and $l = 0.2$. This function corresponds to the operator $\delta$ as shown in section \ref{sec: temporal averaging}.}
 \label{fig: profiles}
\end{figure}

Learning corresponds to a modification of the connectivity simultaneous to the network activity evolution. The result is a coupled system of equations. The learning rate $\eps$ is chosen to be small: $\eps \ll 1$, so that learning can be considered slow compared to the evolution of the activity. The full online learning system is
\begin{equation}
\label{eq: ze learning rule}
\left\{
\begin{array}{c}
 \dot \v = -l\v + \W.S(\v) + \u \\
 \\
\frac{1}{\eps}\dot{\W}_{ij} = \underbrace{\delta[\bar{\v}_i,S(\bar{\v}_j)]}_{\mbox{STDP}} - \underbrace{\sum_k \W_{ik} S(\bar{\v}_k) S(\bar{\v}_j)}_{\mbox{homeostatic plasticity}}
\end{array}
\right.
\end{equation}
where
\begin{equation*}
\bar{\v} = l\v - \W.S(\v)\ast g_l 
\end{equation*}
and
\begin{equation}
\delta[x,y] = \frac{\gamma + l}{2} x (y*g_\gamma) - \frac{\gamma - l}{2} (x*g_\gamma) y
\label{eq: delta}
\end{equation}
where the notation $\ast$ denotes the convolution operator. The function $g_c$ is defined as $g_c:t \mapsto ce^{-c t}H(t)$ with $H$ the Heaviside function and for any positive number $c$ as shown in the left picture of Figure \ref{fig: profiles}. As illustrated in the right picture of Figure \ref{fig: profiles}, the operator $\delta$ roughly corresponds to the classical STDP window \cite{bi1998synaptic} (taking into account a y-axis symmetry corresponding to the symmetric formalism we are using). The constant $\gamma \in \R_+$ is a time constant corresponding to  the width of the STDP window used for learning.

\subsection{Biological mechanisms}
\subsubsection{An input estimate}
The variable $\bar{\v}$ can be seen as a spatio-temporal differential variable which approximates the inputs $\u$. Although unsupervised learning rules are often algebraic combinations of element-wise functions applied to the activity of the network~\cite{gerstner2002spiking}, it is not precisely the case here. Indeed, learning is based on the variable $\bar{\v}$ which corresponds to the subtraction of the temporally integrated synaptic drive $\W.S(\v) \ast g_l$ from the activity of the neurons $l\v$. For each neuron, this variable takes into account the past of all the neurons which are then spatially averaged through the connectivity to be subtracted from the current activity. This gives a differential flavor to this variable which is reminiscent of former learning rules~\cite{bienenstock1982theory, sejnowski1977statistical} for the temporal aspect and~\cite{miller1994role} for the spatial aspect. Note that this variable is not strictly speaking local (i.e. the connection $\W_{ij}$ needs the values 
of $\v_k$ to be updated), yet it is biologically plausible since the term $\big(\W.S(\v)\big)_i$ is accessible for neuron $i$ on its dendritic tree, which is a form of locality in a broader sense.

 \subsubsection{An STDP mechanism}
 The first term $\delta[\bar{\v}_i,S(\bar{\v}_j)]$ in \eqref{eq: ze learning rule} can be related to STDP. The antisymmetric part of this term is responsible for retrieving the drift $\xi$ in equation \eqref{eq: entropy gradient}. The symmetric part (corresponding to Hebbian learning) is responsible for retrieving the second term in \eqref{eq: entropy gradient}. Thus, it captures the causality structure of the inputs which is a task generally attributed to STDP~\cite{sjostrom:2010}. Beyond the simple similarity of functional role, we believe a simplification of this term may shed light on the deep link it has with STDP. The main difference between our setup and STDP is that the former is based on a rate-based dynamics, whereas the latter is based on a spiking dynamics. In a pure spike framework, i.e the activity is a sum of Diracs, the STDP can be seen as this simple learning rule $\dot{\W}_{ij} = \delta[\v_i, \v_j] = \frac{\gamma + l}{2} \v_i (\v_j \ast g_\gamma) - \frac{\gamma - l}{2}(\v_i \ast g_\gamma) \
v_j$. Indeed, the term $\v_i (\v_j \ast g_\gamma)$ is non-null only when the post-synaptic neuron $i$ is firing and then, via the factor $\v_j \ast g_\gamma$, it counts the number of preceding pre-synaptic spikes that might have caused $i$'s excitation and weight them by the decreasing exponential $g_\gamma$. Thus, this term exactly accounts for the positive part of the STDP curve. The negative term $- (\v_i \ast g_\gamma) \v_j$ takes the opposite perspective and accounts for the negative part of the STDP curve. A loose extension of this rule to the case where the activity is smoothly evolving leads to identifying the function $\delta$ to the STDP mechanism for rate-based networks~\cite{galtier2012doctorat, izhikevich2003relating}.

\subsubsection{Homeostatic plasticity}
The second term $- \sum_k \W_{ik} S(\bar{\v}_k) S(\bar{\v}_j)$ in \eqref{eq: ze learning rule} {accounts for what is usually presented as homeostatic plasticity mechanisms.} The previous STDP term seems to be a powerful mechanism to shape the response of the network. However, there is a need of a regulatory process to prevent from uncontrolled growth of the network connectivity~\cite{abbott2000synaptic, turrigiano2004homeostatic, miller1996synaptic}. It has been argued that STDP could be self regulatory~\cite{van2000stable, song2000competitive}, but it is not the case in our framework and an explicit balancing mechanism is necessary to avoid the divergence of the system. This last term {is} the only one with a negative sign and is multiplicative with respect to the connectivity. Thus, according to~\cite{abbott2000synaptic}, it is a reasonable candidate for homeostasis. It has been argued~\cite{turrigiano1998activity, kim2012improved} that homeostatic plasticity might keep the 
relative synaptic weights by dividing the connectivity with a common scaling factor, theoretically preventing from a possible information loss. In contrast to these \textit{ad hoc} re-normalizations often introduced in other learning rules~\cite{miller1994role, oja1982simplified}, our relative entropy minimizing learning rule thus introduces {naturally} an original form of homeostatic plasticity. 

Although we have separated the description of the various terms in \eqref{eq: ze learning rule}, our approach suggests that homeostasis may be seen, not necessarily as a scaling term, but as a constitutive part of a learning principle, deeply entangled \cite{turrigiano1999homeostatic} with the other learning mechanisms.

\subsection{Numerical application}
\begin{figure}[ht!]
 \centering
 \includegraphics[width=1\textwidth]{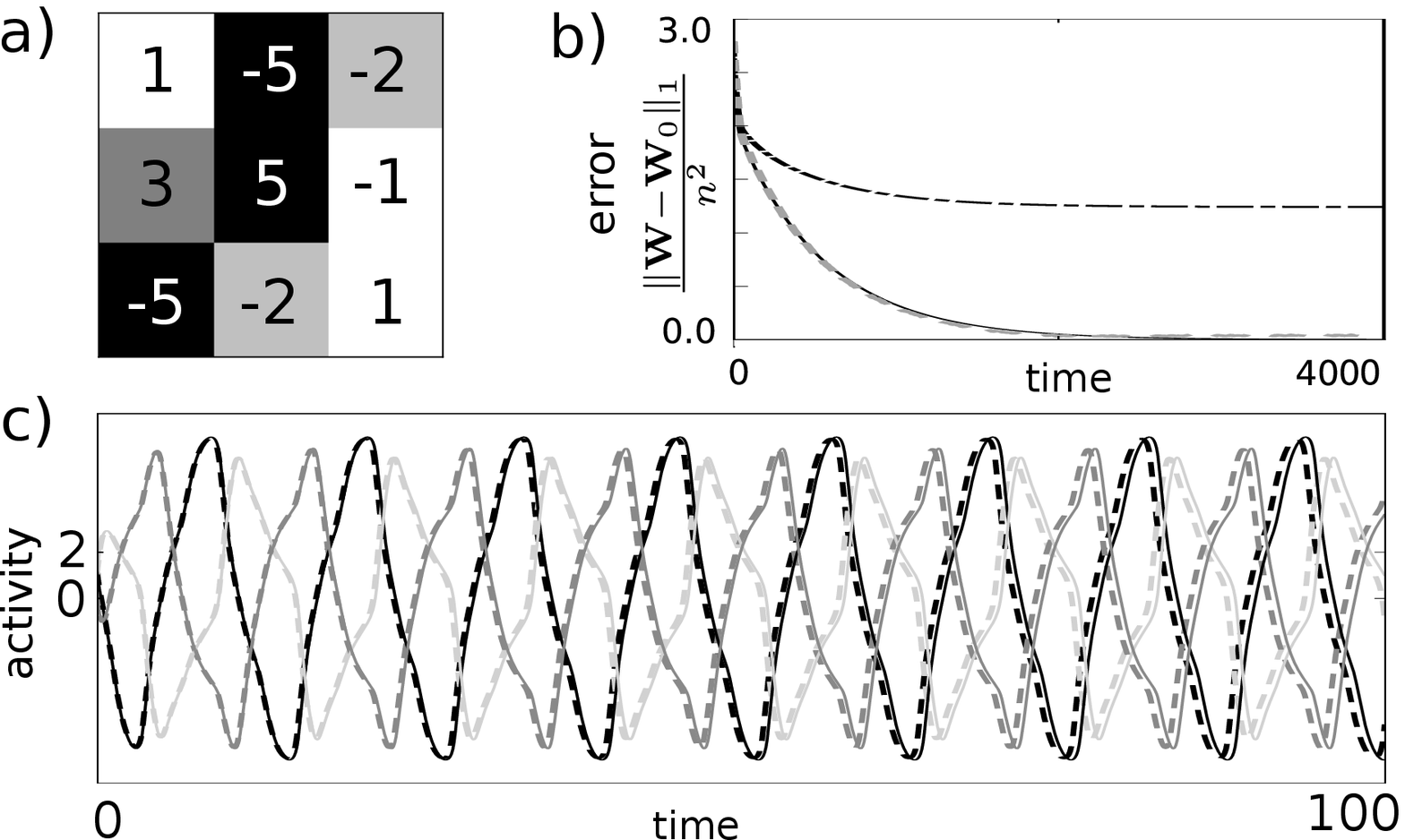}
 \caption{{\bf Retrieving the connectivity} a) Connectivity matrix $\W_0$ of the input network. b) Evolution of the difference between current connectivity and input connectivity through learning. The black dot-dashed curve corresponds to the online learning rule \eqref{eq: ze learning rule} in the homogeneous case, i.e. $l = l_{net} = l_0$ in both equations in \eqref{eq: ze learning rule}. The grey dashed curve corresponds to the online learning system \eqref{eq: ze learning rule} in the hybrid framework, i.e. $l = l_0$ for the learning equation and $l = l_{net} \gg l_0$ for the network equation. For this simulation we chose $l_{net} = 50$. The black plain curve (superposed to the grey dashed curve) corresponds to the batch relative entropy minimization \eqref{eq: entropy gradient}. c) The dashed curves correspond to the activity of the inputs. It is a three dimensional input (i.e. $n = 3$) and the three different grey levels correspond to the different dimensions. The plain curves correspond to the 
simulation of the network (top equation in \eqref{eq: ze learning rule}) post learning, in the hybrid framework, and without inputs. The parameters for these simulations are $l_0 = 1$, $\eps = 0.01$ and $\gamma = 100$.}
 \label{fig: example 1}
\end{figure}

Although the focus of the paper is on theory, we introduce a simple numerical example to illustrate the predictive properties of the biological learning rule. More precisely, we investigate the question of retrieving the connectivity of a neural network based on the observation of the time series of its activity. This is an inverse problem which is a usual challenging topic in computational neuroscience~\cite{friston2003dynamic, galan2008network, potthast2009inverse} since it may give access to large scale effective connectivities simply from the observation of a neuronal activity. Here we address it in an elementary framework. The network generating the activity patterns is referred as input network and evolves according to $\dot \u = -l_0 \u + \W_0.S(\u)$. For this example, the network is made of $n = 3$ neurons and its connectivity $\W_0$ is shown in Figure \ref{fig: example 1}.a). These parameters were chosen so that the activity is periodic as shown by the dashed curves in Figure \ref{fig: 
example 1}.c). Then, we simulate the entire system \eqref{eq: ze learning rule} with a decay constant $l_{net}$ and observe that its connectivity $\W_{net}$ converges to $\W_0$.

As shown in the next section, it is necessary $\u \ast g_l \ast g_\gamma \simeq \u$ in order to approximate accurately the input's activity with the online learning rule \eqref{eq: ze learning rule}. Given that the time constant of the inputs is governed by $l_0$, the previous approximation holds only if $\gamma, l_{net}\gg l_0$. If this assumption is broken, e.g. $l_{net} = l_0$, then the final connectivity matrix is different from $\W_0$, see the black dot-dashed curve in Figure \ref{fig: example 1}.b). Indeed, in this case, the network only learns to replay the slow variation of the inputs.

A method to recover the precise time course of the inputs consists in artificially changing the time constants at different steps of an algorithm described in the following. First, simulate the network equation (top equation in \eqref{eq: ze learning rule}) with a constant $l_{net} \gg l_0$. Yet, the time constant in the learning equation (bottom equation in \eqref{eq: ze learning rule}) is to be kept at $l_0$, thus introducing a hybrid model. In this framework, the connectivity converges exactly to $\W_0$ as shown in the grey dashed curve in Figure \ref{fig: example 1}.b). After learning, simulate the network equation with the learned connectivity and with a time constant switched back to $l_0$. This gives an activity as shown in the plain curves in Figure \ref{fig: example 1}.c).

\section{Link between theoretical and biological learning rules}\label{sec: link}
In this part, we show how the biological learning rule \eqref{eq: ze learning rule} implements the gradient descent based on equation \eqref{eq: entropy gradient}. First, we introduce three technical results which are the mathematical cornerstones of the paper and then combine them to obtain the desired result.

\subsection{Three technical results}
As mentioned previously, equation \eqref{eq: entropy gradient} is not biologically plausible for three main reasons: (i) it requires a direct access to the inputs $\u$, whereas synaptic plasticity mechanisms shall only rely on the network activity $\v$; (ii) it is a batch learning algorithm which requires an access to the entire history of the inputs; (iii) it requires the ability to compute $\xi(\u)$ (equal to the time-derivative of the inputs according to equation \eqref{eq: inputs dynamics}).

On the contrary, the biological learning rule \eqref{eq: ze learning rule} does not have these problems. First, it uses an estimate of the inputs, noted $\bar \v$, which is based on the activity variable only. Second, it is an online learning rule, i.e. it takes input on the fly, and relies on a slow-fast mechanism to temporally average the variables. Third, it approximates the computation of $\xi(\u)$ with a function $\delta$ inspired from STDP convolution. In the following, we mathematically address these three points successively.

\subsubsection{$\bar \v$ is an input estimate}
The online learning rule \eqref{eq: ze learning rule} is expressed by means of the activity of the neural network $\v$ governed by the top equation in \eqref{eq: ze learning rule}. However, to be comparable to the gradient of the relative entropy \eqref{eq: entropy gradient}, we first need to make explicit the dependency on the inputs $\u$. Therefore, we show that the network dynamics induces a simple relation between $\bar{\v}$ and the inputs $\u$: a simple computation in the Fourier domain shows that the convolution between the temporal operator $\frac{d}{dt} + lI_d$ and $g_l$ results in $lI_d$. Applying this result to the neural dynamics leads to reformulating the top equation in \eqref{eq: ze learning rule} as $l \v - \W\cdot S(\v) \ast g_l = \u \ast g_l$. We recognize the definition of the variable $\bar{\v}$ (which was originally defined according to this relation) such that the network's dynamics of the fast equation in \eqref{eq: ze learning rule} is equivalent to
\begin{equation}
\bar{\v} = \u\ast g_l
\label{eq: network dynamics reformulated}
\end{equation}

\subsubsection{Temporal averaging of the learning rule}\label{sec: temporal averaging}
To prove that \eqref{eq: ze learning rule} implements the gradient descent of the relative entropy, we need to use a time-scale separation assumption, enabling the input to reveal its dynamical structure through ergodicity. Indeed, the fact that learning is very slow compared to the activity $\v$ corresponds to the assumption $\eps \ll 1$. In this case, we can apply classical results of periodic averaging~\cite{sanders1985averaging} (see also \cite{galtier2012multiscale} in the context of neural networks) to show that the evolution of $\W$ is well-approximated by
\begin{equation}
 \dot\W = \big[\bar{\v} \ast \Delta_\gamma + l\bar{\v} \ast \Sigma_\gamma - \W. S(\bar{\v})\big]\cdot S(\bar{\v})'
\label{eq: ze learning rule averaged 2}
\end{equation}
where $\Delta_\gamma: t \mapsto \frac{\gamma}{2}\big(g_{\gamma}(-t) - g_{\gamma}(t)\big)$ and $\Sigma_\gamma:t \mapsto \frac{g_{\gamma}(-t) + g_{\gamma}(t)}{2}$. The right picture of Figure \ref{fig: profiles} shows a plot of the function $\Delta_\gamma + l \Sigma_\gamma$.

\subsubsection{STDP as a differential operator}
Here, we prove the two following equalities
\begin{equation}
[\x \ast \Sigma_\gamma ]\cdot \y' = (\x\ast g_\gamma) \cdot (\y \ast g_\gamma)'
\end{equation}
and
\begin{equation}
[\x \ast \Delta_\gamma ]\cdot \y' = (\dot{\x}\ast g_\gamma) \cdot (\y \ast g_\gamma)'
\label{eq: STDP diff op}
\end{equation}
This second equation is the key mathematical mechanism that makes STDP a good approximation of the temporal derivative of the inputs.

Both proofs consist in going in the Fourier domain, where convolutions are turned into multiplications. We use the unitary, ordinary frequency convention for the Fourier transform. Observe that the Fourier transform of $g_\gamma$ is $\hat{g}_\gamma(\xi) = \frac{\gamma}{\gamma + 2i \pi \xi}$. And we define\footnote{\doublespacing This notation is motivated by the following: the convolution with $g_\gamma$ can be written as a matrix multiplication with a continuous Toeplitz matrix $\R^{\R \times \R}$ whose component $ts$ is $g_\gamma(t-s)$. In this framework, the convolution with $g_\gamma'$ corresponds to the multiplication with the transpose of the previous Toeplitz operator.} $g_\gamma':t \mapsto g_\gamma(-t)$ such that $\Delta_\gamma = \frac{\gamma}{2}(g_\gamma' - g_\gamma)$ and $\hat{g_\gamma'}(\xi) = \hat{g}_\gamma(-\xi)$.
\begin{enumerate}
\item Let us show that $[\x \ast \Sigma_\gamma ]\cdot \y' = (\x \ast g_\gamma) \cdot (\y \ast g_\gamma)'$\\
We proceed in two steps:
 \begin{enumerate}
 \item Let us show that $\frac{g_\gamma + g_\gamma'}{2} = g_\gamma \ast g_\gamma'$\\
 The Fourier transform of the convolution $g_\gamma \ast g_\gamma'$ is the product
$\frac{\gamma}{\gamma + 2i\pi\xi}\frac{\gamma}{\gamma - 2i\pi\xi} = \frac{\gamma^2}{\gamma^2 + 4\pi^2 \xi^2}$. From the usual Fourier tables we observe that the right hand side is the Fourier transform of $t \mapsto \frac{\gamma}{2} e^{-\gamma |t|}$. This immediately leads to the result.
\item Let us show that $(\x \ast g_\gamma') \cdot \y' = \x  \cdot (\y \ast g_\gamma)'$\\
Compute
\begin{multline*}
 \{(\x \ast g_\gamma') \cdot \y'\}_{ij} = \int_{-\infty}^{\infty} \big(\x_i \ast g_\gamma'\big)(s) \y_j(s) ds
 = \int_{-\infty}^{\infty} \int_{-\infty}^{+\infty} \x_i(t) g_\gamma'(s-t) \y_j(s) ds dt\\
 = \int_{-\infty}^{\infty} \int_{-\infty}^{+\infty} \x_i(t) g_\gamma(t-s) \y_j(s) ds dt = \{\x \cdot (\y \ast g_\gamma)'\}_{ij}
\end{multline*}
\end{enumerate}
Using the result (a) and applying result (b) to $\x\ast g_\gamma$ and $\y$ leads to the result $[\x \ast \Sigma_\gamma ]\cdot \y' = (\x \ast g_\gamma) \cdot (\y \ast g_\gamma)'$.

 \item To prove equation \eqref{eq: STDP diff op}, let us first show that $\x \ast \Delta_\gamma = \dot \x \ast \frac{g_\gamma + g_\gamma'}{2}$.\\
The Fourier transform of the convolution $\x \ast \Delta_\gamma$ is the product $\hat{\x}\ \hat{\Delta}_\gamma$. Besides, 
\begin{multline*}
 \frac{2}{\gamma}\hat{\Delta}_\gamma = \widehat{g_{\gamma}' - g_{\gamma}}(\xi) = \frac{\gamma}{\gamma - 2i \pi \xi} - \frac{\gamma}{\gamma + 2i \pi \xi} = 2i\pi\xi\frac{2\gamma}{\gamma^2 + 4 \pi^2 \xi^2}\\
=\frac{2i\pi\xi}{\gamma}\Big(\frac{\gamma}{\gamma + 2i \pi \xi} + \frac{\gamma}{\gamma - 2i \pi \xi}\Big) = \frac{2i\pi\xi}{\gamma}\big(\widehat{g_{\gamma}' + g_{\gamma}}(\xi)\big)
\end{multline*}
Because $\hat{\frac{d\x}{dt}}(\xi) = \hat{\x}\ 2i\pi\xi$, taking the inverse Fourier transform of $\hat{\x}\ 2i\pi\xi \widehat{\frac{g_{\gamma}' + g_{\gamma}}{2}}(\xi)$ gives the intermediary result $\x \ast \Delta_\gamma = \dot \x \ast \frac{g_\gamma + g_\gamma'}{2}$.

Using it and applying the first result to $\dot{\x}$ and $\y$ leads to equation \eqref{eq: STDP diff op}.
\end{enumerate}

\subsection{Main result}
We prove here that the biological learning rule \eqref{eq: ze learning rule} implements the gradient descent based on \eqref{eq: entropy gradient}.

The temporally averaged version of the biological learning rule is given by \eqref{eq: ze learning rule averaged 2}. We simultaneously use equations \eqref{eq: network dynamics reformulated} and \eqref{eq: STDP diff op} on this formula to show that the solution of the biological learning rule is well approximated by
\begin{multline*}
 \dot\W = \big[\xi(\u \ast g_l)\ast g_\gamma\big]\cdot \big[S(\u \ast g_l) \ast g_\gamma\big]' + l (\u \ast g_l \ast g_\gamma)\cdot \big(S(\u \ast g_l)\ast g_\gamma\big)'\\
 - \W\cdot S(\u \ast g_l)\cdot S(\u \ast g_l)'
\end{multline*}

If $\u$ is slow enough, i.e. $\u \ast g_l \ast g_\gamma \simeq \u$, then this equation is precisely the gradient descent of $H_{\u}$. If $\u$ is too fast, then the network will only learn to predict the slow variations of the inputs. Some fast-varying information is lost in the averaging process, mainly because the network equation in \eqref{eq: ze learning rule} acts as a relaxation equation which {filters} the activity.  Note that this loss of information is not necessarily a problem for the brain, since it may be extracting and treating information at different time scales~\cite{kiebel2008hierarchy}. Actually, the choice of the parameter $l$ specifies the time scale under which the inputs are observed.

\subsection{Stability}
Both theoretical and biological learning rules are assured to make the connectivity converge to an equilibrium point whatever the initial condition. In particular, this method does not suffer from the bifurcation issues often encountered in recurrent neural network learning \cite{doya1992bifurcations}. In most cases, problems arise when learning leads the network activity to a bifurcation. The bifurcation suddenly changes the value of the quantity being minimized and this intricate coupling leads to very slow convergence rates. The reason why this method has such an unproblematic converging property is because the relative entropy (as opposed to other energy functions traditionally used) is independent from the network activity $\v$. Besides, for any network pattern $\v$ equation \eqref{eq: network dynamics reformulated} ensures that $\bar \v$ is always equal to the convolved inputs. These two arguments break the pathologic coupling preventing from getting interesting convergence results.

From a mathematical perspective, the Krasovskii-Lasalle invariance principle~\cite{khalil1992nonlinear} ensures that the theoretical learning rule $\dot\W=-\nabla_\W H_\u$ converges to a equilibrium point. Indeed, the relative entropy acts as an energy or Lyapunov function. If $S(\u).S(\u)'$ is invertible then the equilibrium point is unique and defined by equation \eqref{eq: equilbrium connectivity}. If it is not invertible then the equilibrium depends on the initial condition.

\noindent If the inputs are slow enough, it has been shown that the biological learning is well approximated by the theoretical gradient descent. Therefore, the former exhibits the same stable converging behavior as the latter. Thus, the biological learning rule \eqref{eq: ze learning rule} is stable. In practice, we have never encountered a diverging situation even when the inputs are not slow enough.

\section{Discussion and conclusion}
We have shown that a biological learning rule shapes a network into a predictor of its stimuli. This learning rule is made of a combination of a STDP mechanism and homeostatic plasticity. After learning the network would spontaneously predict and replay the stimuli it has previously been exposed to. This was achieved by showing that this learning rule minimizes a quantity analogous to the relative entropy (or Kullback-Leibler divergence) between the stimuli and the network activity as for other well-known algorithms \cite{ackley1985learning, hinton2009deep}.

We believe this letter brings interesting arguments in the debate concerning the functional role of STDP. We have shown that the antisymmetric part of STDP can be seen as a differential operator. When its effect is moderated by an appropriate scaling term, we argue that it could correspond to a generic mechanism shaping neural networks into predictive units. This idea is not new, but this letter may give it a stronger theoretical basis.

This study also gives central importance to the time constants characterizing the network. Indeed, the fact the biological learning rule \eqref{eq: ze learning rule} implements a gradient descent is rigorously true for slow inputs. Inputs are slow if they are significantly slower than both the time window of the STDP and the decay constant of the network. This means that sub-networks in the brain could select the frequency of the information they are predicting. Given that the brain may process information at different time scales \cite{kiebel2008hierarchy}, this may be an interesting feature. Besides, note that the proposed biological learning rule \eqref{eq: ze learning rule} is partly characterized by the activity decay parameter $l$. This is surprising because it links the intrinsic dynamics of the neurons to the learning processes corresponding to the synapse. Therefore, it may provide a direction to experimentally check this theory: is the symmetric part of STDP (i.e. Hebbian learning) stronger 
between faster neurons?

One of the characteristic features of this research is the combination of rate-based networks and the concept of spike timing dependent plasticity. Obviously, this made impossible to consider the rigorous definition of STDP. However, we have argued that the function $\delta$ in \eqref{eq: delta} is an alternative formulation which is equivalent to the others in the spiking framework and which can be trivially extended to analog networks. Besides, it does capture the functional mechanism of the STDP in the rate-based framework: when the activation of a neuron precedes (resp. follows) the activation of another the strength of the synapse from the former to the latter is increased (resp. decreased). Finally, we believe the theory presented in this paper gives support to this rate-based STDP since it appears to fill a gap between machine learning and biology by implementing a differential operator.

This approach can be applied to other forms of network equations than \eqref{eq: voltage based} such as the Wilson-Cowan or Kuramoto models, leading to different learning rules. In this perspective, learning can be seen as a projection of a given arbitrary dynamical system to a versatile neuronal network model, and the learning rule will depend on the chosen model. However, we shall remark that any network equation with an additive structure -- intrinsic dynamics + communication with other neurons -- as in \eqref{eq: voltage based} will lead to a similar structure for the learning rule, with terms that may share similar biological interpretations as developed above. A special case is the linear network $\dot \v = -l\v + \W.\v$ for which various statistical methods to estimate the connectivity matrix have been applied e.g. in climate modeling~\cite{penland1995optimal}, gene regulatory networks~\cite{yeung2002reverse} and spontaneous neuronal activity~\cite{galan2008network}. In this simpler case, the 
biological learning rule in equation \eqref{eq: ze learning rule} would be $\frac{1}{\eps}\dot{\W}_{ij} = \delta[\bar{\v}_i,\bar{\v}_j] - \sum_k \W_{ik} \bar{\v}_k \bar{\v}_j$ with $\bar{\v} = l\v - \W.\v\ast g_l$. The method developed in this article can be used to extend the \textit{inverse modeling} approach previously developed in the linear case to models with non-linear interactions.

One of the main restrictions of this approach is that the dimensionality of the stimuli and that of the network have to be identical. Therefore, as such, the accuracy of this biological mechanism does not match the state of the art machine learning algorithms.This is not necessarily a big issue since a high dimensional projections of the inputs may be used as preprocessing. From a biological viewpoint and taking the example of vision, this would correspond to the retino-cortical pathway which is not one to one and very redundant. But this limitation also puts forward the necessity to study a network containing hidden neurons in a similar fashion. Ongoing research is revealing that the mathematical formalism is well suited to extend this approach to the field of reservoir computing \cite{jaeger2004harnessing}.

\section*{Acknowledgments}
The authors thank Herbert Jaeger for his helpful comments on the manuscript.

MNG was partially funded by the Amarsi project (FP7-ICT \#24833), the ERC advanced grant NerVi \#227747, the r\'egion PACA, France and the IP project BrainScaleS \#269921.

\bibliographystyle{apalike}

\begin{thebibliography}{}

\bibitem[Abbott and Nelson, 2000]{abbott2000synaptic}
Abbott, L. and Nelson, S. (2000).
\newblock Synaptic plasticity: taming the beast.
\newblock {\em Nature neuroscience}, 3:1178--1183.

\bibitem[Ackley et~al., 1985]{ackley1985learning}
Ackley, D., Hinton, G., and Sejnowski, T. (1985).
\newblock A learning algorithm for boltzmann machines.
\newblock {\em Cognitive science}, 9(1):147--169.

\bibitem[Bar, 2009]{bar2009predictions}
Bar, M. (2009).
\newblock Predictions: a universal principle in the operation of the human
  brain.
\newblock {\em Philosophical Transactions of the Royal Society B: Biological
  Sciences}, 364(1521):1181--1182.

\bibitem[Berkes et~al., 2011]{berkes2011spontaneous}
Berkes, P., Orb{\'a}n, G., Lengyel, M., and Fiser, J. (2011).
\newblock Spontaneous cortical activity reveals hallmarks of an optimal
  internal model of the environment.
\newblock {\em Science}, 331(6013):83.

\bibitem[Bi and Poo, 1998]{bi1998synaptic}
Bi, G. and Poo, M. (1998).
\newblock Synaptic modifications in cultured hippocampal neurons: dependence on
  spike timing, synaptic strength, and postsynaptic cell type.
\newblock {\em The Journal of Neuroscience}, 18(24):10464--10472.

\bibitem[Bienenstock et~al., 1982]{bienenstock1982theory}
Bienenstock, E., Cooper, L., and Munro, P. (1982).
\newblock Theory for the development of neuron selectivity: orientation
  specificity and binocular interaction in visual cortex.
\newblock {\em The Journal of Neuroscience}, 2(1):32--48.

\bibitem[Bitzer and Kiebel, 2012]{bitzer2012recognizing}
Bitzer, S. and Kiebel, S. (2012).
\newblock Recognizing recurrent neural networks (rrnn): Bayesian inference for
  recurrent neural networks.
\newblock {\em Biological cybernetics}, pages 1--17.

\bibitem[Caporale and Dan, 2008]{caporale2008spike}
Caporale, N. and Dan, Y. (2008).
\newblock Spike timing-dependent plasticity: a hebbian learning rule.
\newblock {\em Annu. Rev. Neurosci.}, 31:25--46.

\bibitem[Clark, 2012]{clark2012whatever}
Clark, A. (2012).
\newblock Whatever next? predictive brains, situated agents, and the future of
  cognitive science.
\newblock {\em Behav. Brain Sci}.

\bibitem[Dayan et~al., 1995]{dayan1995helmholtz}
Dayan, P., Hinton, G., Neal, R., and Zemel, R. (1995).
\newblock The helmholtz machine.
\newblock {\em Neural computation}, 7(5):889--904.

\bibitem[Deneve, 2008]{deneve2008bayesian}
Deneve, S. (2008).
\newblock Bayesian spiking neurons i: inference.
\newblock {\em Neural computation}, 20(1):91--117.

\bibitem[Doya, 1992]{doya1992bifurcations}
Doya, K. (1992).
\newblock Bifurcations in the learning of recurrent neural networks.
\newblock In {\em Circuits and Systems, 1992. ISCAS'92. Proceedings., 1992 IEEE
  International Symposium on}, volume~6, pages 2777--2780. IEEE.

\bibitem[Friston, 2010]{friston2010free}
Friston, K. (2010).
\newblock The free-energy principle: a unified brain theory?
\newblock {\em Nature Reviews Neuroscience}, 11(2):127--138.

\bibitem[Friston et~al., 2003]{friston2003dynamic}
Friston, K., Harrison, L., and Penny, W. (2003).
\newblock Dynamic causal modelling.
\newblock {\em Neuroimage}, 19(4):1273--1302.

\bibitem[Gal{\'a}n, 2008]{galan2008network}
Gal{\'a}n, R. (2008).
\newblock On how network architecture determines the dominant patterns of
  spontaneous neural activity.
\newblock {\em PLoS One}, 3(5):e2148.

\bibitem[Galtier, 2012]{galtier2012doctorat}
Galtier, M. (2012).
\newblock {\em A mathematical approach to unsupervised learning in recurrent
  neural networks}.
\newblock PhD thesis, Mines Paristech / INRIA.

\bibitem[Galtier and Wainrib, 2012]{galtier2012multiscale}
Galtier, M. and Wainrib, G. (2012).
\newblock Multiscale analysis of slow-fast neuronal learning models with noise.
\newblock {\em The Journal of Mathematical Neuroscience}, 2(1):13.

\bibitem[George and Hawkins, 2009]{george2009towards}
George, D. and Hawkins, J. (2009).
\newblock Towards a mathematical theory of cortical micro-circuits.
\newblock {\em PLoS computational biology}, 5(10):e1000532.

\bibitem[Gerstner and Kistler, 2002]{gerstner2002spiking}
Gerstner, W. and Kistler, W. (2002).
\newblock {\em Spiking neuron models: Single neurons, populations, plasticity}.
\newblock Cambridge Univ Pr.

\bibitem[Gerstner et~al., 1993]{gerstner1993spikes}
Gerstner, W., Ritz, R., and Van~Hemmen, J. (1993).
\newblock Why spikes? hebbian learning and retrieval of time-resolved
  excitation patterns.
\newblock {\em Biological cybernetics}, 69(5):503--515.

\bibitem[Gilbert and Wilson, 2007]{gilbert2007prospection}
Gilbert, D. and Wilson, T. (2007).
\newblock Prospection: experiencing the future.
\newblock {\em Science}, 317(5843):1351--1354.

\bibitem[Hennequin et~al., 2010]{hennequin2010stdp}
Hennequin, G., Gerstner, W., and Pfister, J. (2010).
\newblock Stdp in adaptive neurons gives close-to-optimal information
  transmission.
\newblock {\em Frontiers in Computational Neuroscience}, 4.

\bibitem[Hinton, 2009]{hinton2009deep}
Hinton, G.~E. (2009).
\newblock Deep belief networks.
\newblock {\em Scholarpedia}, 4(4):5947.

\bibitem[Izhikevich and Desai, 2003]{izhikevich2003relating}
Izhikevich, E. and Desai, N. (2003).
\newblock Relating stdp to bcm.
\newblock {\em Neural Computation}, 15(7):1511--1523.

\bibitem[Jaeger and Haas, 2004]{jaeger2004harnessing}
Jaeger, H. and Haas, H. (2004).
\newblock Harnessing nonlinearity: Predicting chaotic systems and saving energy
  in wireless communication.
\newblock {\em Science}, 304(5667):78--80.

\bibitem[Kempter et~al., 1999]{kempter1999hebbian}
Kempter, R., Gerstner, W., and Van~Hemmen, J. (1999).
\newblock Hebbian learning and spiking neurons.
\newblock {\em Physical Review E}, 59(4):4498.

\bibitem[Kenet et~al., 2003]{kenet2003spontaneously}
Kenet, T., Bibitchkov, D., Tsodyks, M., Grinvald, A., and Arieli, A. (2003).
\newblock Spontaneously emerging cortical representations of visual attributes.
\newblock {\em Nature}, 425(6961):954--956.

\bibitem[Khalil and Grizzle, 1992]{khalil1992nonlinear}
Khalil, H. and Grizzle, J. (1992).
\newblock {\em Nonlinear systems}.
\newblock Macmillan Publishing Company New York.

\bibitem[Kiebel et~al., 2008]{kiebel2008hierarchy}
Kiebel, S., Daunizeau, J., and Friston, K. (2008).
\newblock A hierarchy of time-scales and the brain.
\newblock {\em PLoS computational biology}, 4(11):e1000209.

\bibitem[Kim et~al., 2012]{kim2012improved}
Kim, J., Tsien, R., and Alger, B. (2012).
\newblock An improved test for detecting multiplicative homeostatic synaptic
  scaling.
\newblock {\em PloS one}, 7(5):e37364.

\bibitem[Lazar et~al., 2007]{lazar2007}
Lazar, A., Pipa, G., and Triesch, J. (2007).
\newblock Fading memory and time series prediction in recurrent networks with
  different forms of plasticity.
\newblock {\em Neural Networks}, 20(3):312--322.

\bibitem[Lazar et~al., 2009]{lazar2009sorn}
Lazar, A., Pipa, G., and Triesch, J. (2009).
\newblock Sorn: a self-organizing recurrent neural network.
\newblock {\em Frontiers in computational neuroscience}, 3.

\bibitem[Lukosevicius and Jaeger, 2009]{lukovsevivcius2009reservoir}
Lukosevicius, M. and Jaeger, H. (2009).
\newblock Reservoir computing approaches to recurrent neural network training.
\newblock {\em Computer Science Review}, 3(3):127--149.

\bibitem[Mandic and Chambers, 2001]{mandic2001recurrent}
Mandic, D. and Chambers, J. (2001).
\newblock {\em Recurrent neural networks for prediction: Learning algorithms,
  architectures and stability}.
\newblock John Wiley \& Sons, Inc.

\bibitem[Markram et~al., 1997]{markram1997regulation}
Markram, H., L{\"u}bke, J., Frotscher, M., and Sakmann, B. (1997).
\newblock Regulation of synaptic efficacy by coincidence of postsynaptic aps
  and epsps.
\newblock {\em Science}, 275(5297):213--215.

\bibitem[Miller, 1996]{miller1996synaptic}
Miller, K. (1996).
\newblock Synaptic economics: Competition and cooperation in correlation-based
  synaptic plasticity.
\newblock {\em Neuron}, 17:371--374.

\bibitem[Miller and MacKay, 1994]{miller1994role}
Miller, K. and MacKay, D. (1994).
\newblock The role of constraints in hebbian learning.
\newblock {\em Neural Computation}, 6(1):100--126.

\bibitem[Oja, 1982]{oja1982simplified}
Oja, E. (1982).
\newblock Simplified neuron model as a principal component analyzer.
\newblock {\em Journal of mathematical biology}, 15(3):267--273.

\bibitem[Pearlmutter, 1995]{pearlmutter1995gradient}
Pearlmutter, B. (1995).
\newblock Gradient calculations for dynamic recurrent neural networks: A
  survey.
\newblock {\em Neural Networks, IEEE Transactions on}, 6(5):1212--1228.

\bibitem[Penland and Sardeshmukh, 1995]{penland1995optimal}
Penland, C. and Sardeshmukh, P. (1995).
\newblock The optimal growth of tropical sea surface temperature anomalies.
\newblock {\em Journal of climate}, 8(8):1999--2024.

\bibitem[Pfister and Gerstner, 2006]{pfister2006triplets}
Pfister, J. and Gerstner, W. (2006).
\newblock Triplets of spikes in a model of spike timing-dependent plasticity.
\newblock {\em The Journal of neuroscience}, 26(38):9673--9682.

\bibitem[Potthast and beim Graben, 2009]{potthast2009inverse}
Potthast, R. and beim Graben, P. (2009).
\newblock Inverse problems in neural field theory.
\newblock {\em SIAM Journal on Applied Dynamical Systems}, 8:1405.

\bibitem[Rao et~al., 1999]{rao1999predictive}
Rao, R., Ballard, D., et~al. (1999).
\newblock Predictive coding in the visual cortex: a functional interpretation
  of some extra-classical receptive-field effects.
\newblock {\em Nature neuroscience}, 2:79--87.

\bibitem[Rao and Sejnowski, 2001]{rao2001spike}
Rao, R. and Sejnowski, T. (2001).
\newblock Spike-timing-dependent hebbian plasticity as temporal difference
  learning.
\newblock {\em Neural computation}, 13(10):2221--2237.

\bibitem[Rosenblatt, 1958]{rosenblatt1958perceptron}
Rosenblatt, F. (1958).
\newblock The perceptron: a probabilistic model for information storage and
  organization in the brain.
\newblock {\em Psychological review}, 65(6):386.

\bibitem[Sanders and Verhulst, 1985]{sanders1985averaging}
Sanders, J. and Verhulst, F. (1985).
\newblock {\em Averaging methods in nonlinear dynamical systems}, volume~59.
\newblock Springer.

\bibitem[Schacter et~al., 2008]{schacter2008episodic}
Schacter, D., Addis, D., and Buckner, R. (2008).
\newblock Episodic simulation of future events.
\newblock {\em Annals of the New York Academy of Sciences}, 1124(1):39--60.

\bibitem[Sejnowski, 1977]{sejnowski1977statistical}
Sejnowski, T. (1977).
\newblock Statistical constraints on synaptic plasticity.
\newblock {\em Journal of theoretical biology}, 69(2):385.

\bibitem[Sj\"ostr\"om and Gerstner, 2010]{sjostrom:2010}
Sj\"ostr\"om, J. and Gerstner, W. (2010).
\newblock Spike-timing dependent plasticity.
\newblock 5(2):1362.

\bibitem[Song et~al., 2000]{song2000competitive}
Song, S., Miller, K., and Abbott, L. (2000).
\newblock Competitive hebbian learning through spike-timing-dependent synaptic
  plasticity.
\newblock {\em Nature neuroscience}, 3:919--926.

\bibitem[Sprekeler et~al., 2007]{sprekeler2007slowness}
Sprekeler, H., Michaelis, C., and Wiskott, L. (2007).
\newblock Slowness: An objective for spike-timing--dependent plasticity?
\newblock {\em PLoS Computational Biology}, 3(6):e112.

\bibitem[Sussillo and Abbott, 2009]{sussillo2009generating}
Sussillo, D. and Abbott, L. (2009).
\newblock Generating coherent patterns of activity from chaotic neural
  networks.
\newblock {\em Neuron}, 63(4):544--557.

\bibitem[Turrigiano, 1999]{turrigiano1999homeostatic}
Turrigiano, G. (1999).
\newblock Homeostatic plasticity in neuronal networks: the more things change,
  the more they stay the same.
\newblock {\em Trends in neurosciences}, 22(5):221--227.

\bibitem[Turrigiano et~al., 1998]{turrigiano1998activity}
Turrigiano, G., Leslie, K., Desai, N., Rutherford, L., and Nelson, S. (1998).
\newblock Activity-dependent scaling of quantal amplitude in neocortical
  neurons.
\newblock {\em NATURE}, 391:893.

\bibitem[Turrigiano and Nelson, 2004]{turrigiano2004homeostatic}
Turrigiano, G. and Nelson, S. (2004).
\newblock Homeostatic plasticity in the developing nervous system.
\newblock {\em Nature Reviews Neuroscience}, 5(2):97--107.

\bibitem[Van~Rossum et~al., 2000]{van2000stable}
Van~Rossum, M., Bi, G., and Turrigiano, G. (2000).
\newblock Stable hebbian learning from spike timing-dependent plasticity.
\newblock {\em The Journal of Neuroscience}, 20(23):8812--8821.

\bibitem[Williams and Zipser, 1989]{williams1989learning}
Williams, R. and Zipser, D. (1989).
\newblock A learning algorithm for continually running fully recurrent neural
  networks.
\newblock {\em Neural computation}, 1(2):270--280.

\bibitem[Williams and Zipser, 1995]{williams1995gradient}
Williams, R. and Zipser, D. (1995).
\newblock Gradient-based learning algorithms for recurrent networks and their
  computational complexity.
\newblock {\em Back-propagation: Theory, architectures and applications}, pages
  433--486.

\bibitem[Yeung et~al., 2002]{yeung2002reverse}
Yeung, M., Tegn{\'e}r, J., and Collins, J. (2002).
\newblock Reverse engineering gene networks using singular value decomposition
  and robust regression.
\newblock {\em Proceedings of the National Academy of Sciences}, 99(9):6163.

\bibitem[Yoshioka et~al., 2007]{yoshioka2007spatiotemporal}
Yoshioka, M., Scarpetta, S., and Marinaro, M. (2007).
\newblock Spatiotemporal learning in analog neural networks using
  spike-timing-dependent synaptic plasticity.
\newblock {\em Physical Review E}, 75(5):051917.

\bibitem[Zheng et~al., 2013]{zheng2013network}
Zheng, P., Dimitrakakis, C., and Triesch, J. (2013).
\newblock Network self-organization explains the statistics and dynamics of
  synaptic connection strengths in cortex.
\newblock {\em PLoS Computational Biology}, 9(1).

\end{thebibliography}

\end{document}